\documentclass[12pt,preprint]{aastex}

\lefthead{Antoniou et al.}  \righthead{SFH and XRB populations}

\def\ergcm2s{~erg cm$^{-2}$ s$^{-1}$ } 
 
\def\ergs{~erg s$^{-1}$}             
\def\lunit{~erg s$^{-1}$}

\def\etal{et al.~}              

\def\lunit{~erg s$^{-1}$} 
\def\kms{~km s$^{-1}$}   

\def\msun{~M$_{\odot}$\,}

\def\zsun{Z$_{\odot}$}

\def\deg{$^{\circ}$}

\def\chandra{{\it Chandra~}}
\def\chandrasimple{{\it Chandra}}
\def\xmm{{\it XMM-Newton~}}
\def\xmmsimple{{\it XMM-Newton}}
\def\x2{$\chi^{2}$}

\def\E(B-V){{\it E(B-V)}}

\makeatother

\begin{document}
\title{Star formation history and X-ray binary populations: the case
of the Small Magellanic Cloud}
\author{V. Antoniou\altaffilmark{1,2,3}, A. Zezas\altaffilmark{1,2,4}, D. Hatzidimitriou\altaffilmark{4,5}, V. Kalogera\altaffilmark{6}}
\altaffiltext{1}{Harvard-Smithsonian Center for Astrophysics, 60 Garden Street, Cambridge, MA 02138, USA; vantoniou@head.cfa.harvard.edu}
\altaffiltext{2}{Physics Department, University of Crete, P.O. Box 2208, GR-710 03, Heraklion, Crete, Greece}
\altaffiltext{3}{Department of Physics and Astronomy, Iowa State University, 12 Physics Hall, Ames, IA 50011, USA}
\altaffiltext{4}{IESL/Foundation for Research and Technology-Hellas, P.O. Box 1527, GR-711 10 Heraklion, Crete, Greece}
\altaffiltext{5}{Section of Astrophysics, Astronomy and Mechanics, Department of Physics, University of Athens, Panepistimiopolis, GR-157 84 Zografos, Athens, Greece}
\altaffiltext{6}{Department of Physics and Astronomy, Northwestern University, 2145 Sheridan Road, Evanston, IL 60208, USA}

\begin{abstract}
Using \chandrasimple, \xmm and optical photometric catalogs we study
the young X-ray binary (XRB) populations of the Small Magellanic
Cloud. We find that the Be/X-ray binaries (Be-XRBs) are observed in
regions with star formation rate bursts $\sim$25-60 Myr ago. The
similarity of this age with the age of maximum occurrence of the Be
phenomenon ($\sim$40 Myr) indicates that the presence of a
circumstellar decretion disk plays a significant role in the number of
observed XRBs in the 10-100 Myr age range. We also find that regions
with strong but more recent star formation (e.g., the Wing) are deficient in
Be-XRBs. By correlating the number of observed Be-XRBs with the
formation rate of their parent populations, we measure a Be-XRB
production rate of $\sim$1 system per $3\times10^{-3}$\msun/yr. Finally, we
use the strong localization of the Be-XRB systems in order to set
limits on the kicks imparted on the neutron star during the supernova explosion. 
\end{abstract}


\section{Introduction}\label{intro}
Nearby star-forming galaxies offer a unique environment to study the
young ($<$100 Myr) X-ray binary (XRB) populations. One of the best
cases is the Small Magellanic Cloud (SMC), which at $\sim$60 kpc is our second nearest
star-forming galaxy (Hilditch \etal 2005). Its proximity, well mapped extinction (Zaritsky \etal
2002), moderate Galactic foreground absorption
($\rm{N_{H}}\simeq6\times10^{20} cm^{-2}$; Dickey \& Lockman 1990),
small line-of-sight depth of its young, central stellar populations
($<$10 kpc; Crowl \etal 2001; Harries \etal 2003),
and its well-determined recent star formation history (SFH; Harris \& Zaritsky
2004 [HZ04]) make the SMC the ideal environment for
directly studying the link between XRB populations and star formation (SF). Furthermore, the wealth of multi-wavelength data allows
us to clarify the X-ray sources and obtain an even more precise
picture of their population.

Several studies have compared the number of Be/X-ray binaries (Be-XRBs) in the Magellanic
Clouds and the Galaxy (e.g., Majid \etal 2004, Haberl \& Pietsch 2004, Coe \etal 2005),
concluding that the SMC hosts an unusually large number of these systems.
There is only one identified supergiant XRB located in the SMC
Wing (SMC X-1; Webster \etal 1972) in a population of $\sim$100 High-Mass XRBs (HMXBs;
e.g., Liu \etal 2005, Antoniou \etal 2009b [Paper II]). However, only few of those HMXBs have determined spectral types (e.g., out of the 92 listed in Liu \etal 2005, 53 are cited as Be-XRBs but only 19 have been confirmed spectroscopically). Later works by Antoniou \etal (2009a [Paper I]), Haberl \etal (2008), McBride \etal (2008), Shtykovskiy \& Gilfanov (2005), Coe \etal (2005) and others have increased the number of known Be-XRBs to 67 to date. Nevertheless, this overabundance can be
partly explained by the enhanced SMC SFH $\sim$40 Myr
ago (e.g., Majid \etal 2004, Shtykovskiy \& Gilfanov 2007 [SG07]). However, Antoniou \etal (2009b) show that even after
accounting for the difference in the star formation rate (SFR) between the SMC and the Galaxy, the SMC hosts $\sim$1.5 times more Be-XRBs than the Galaxy
down to a limiting luminosity of ${\rm L_{X}\geq10^{34}}$\ergs. This
residual excess can be explained by the different
metallicity of these galaxies, as justified by population synthesis
models (Dray 2006) and recent observations of Be stars (e.g., Wisniewski \&
Bjorkman 2006, Martayan \etal 2007). The work of Antoniou \etal (2009b) also indicated 
spatial variations of the Be-XRB populations within the SMC Bar, which
could be evidence for small supernova (SN) kicks.

The SMC Bar hosts stellar populations with ages $<$100 Myr [HZ04] and the vast majority
of the SMC pulsars (Galache \etal 2008). [SG07] found that the age distribution of the HMXBs peaks at
$\sim$20-50 Myr after the SF event, while McSwain and Gies (2005)
observed a strong evolution in the fraction of Be stars with age up to
100 Myr, with a maximum at $\sim$25-80 Myr. These results motivated us
to investigate the connection between the
spatially resolved SFH in and around the SMC Bar and the number and spatial distribution of the XRBs. In this study, we use the results from our \chandra survey of the
central, most actively star forming, SMC Bar (A. Zezas \etal 2010, in
preparation; Papers I, II), and data from our \xmm survey of the outer SMC
regions which host young and intermediate age stellar populations
($\sim$10-500 Myr; [HZ04]).

\section{X-ray observations and data analysis}
Using the ACIS-I detector on board \chandra we observed five fields in
the central part of the SMC (the so called SMC Bar), with typical
exposure times of 8-12 ks. These observations yielded a total of 158
sources, down to a limiting luminosity of $\sim4\times10^{33}$\lunit
(0.5-7.0 keV band), reaching the luminosity range of quiescent
HMXBs (typically ${\rm L_{X}\sim10^{33}-10^{35}}$\lunit\,; van Paradijs \& McClintock 1995). The analysis of the data, the source-list and their X-ray luminosity functions are presented in A. Zezas \etal (2010, in preparation), while their optical counterparts and resulting classification are given in Papers I, II.

Our \xmm survey consists of five observations in the outer SMC Bar, performed with the three EPIC (MOS1, MOS2, and PN) detectors in 
full frame mode. One
of these fields was affected by high background flares and it is not
included in this work. The data were analyzed with the \xmm Science Analysis System (SAS)
 version 7.0.0. After processing the raw data with the {\it epchain}
and {\it emchain} tasks, we filtered any bad columns/pixels and high
background flares (excluding times when the total count rate deviated
more than $3\sigma$ from the mean), resulting in 5-18 ks net exposures. We only kept events of patterns 0-4 for the PN and 0-12 for the MOS
detectors. Source detection was performed simultaneously in five energy
bands (0.2-0.5, 0.5-1.0, 1.0-2.0, 2.0-4.5, and
4.5-12.0 keV), and the three EPIC detectors with the maximum likelihood
method (threshold set to 7) of the {\it edetect\_chain} task. The
detected sources were visually inspected for spurious detections. We
detected 186 sources down to a limiting luminosity of $\sim3.5\times10^{33}$\lunit\, (0.2-12 keV), out of which 4-8 sources are expected to be spurious based on the calibration of Watson \etal (2009).

In Table \ref{tableSFH}, we give the ID and the coordinates of the
X-ray fields, along with the properties of the
dominant SF event in each field (see \S \ref{SFHfields}).

\subsection{X-ray source classification}\label{class}
New
HMXBs and candidate Be-XRBs are identified based on their X-ray and optical properties. Hardness ratios between the soft (0.5-1.0 keV), medium
(1.0-2.0 keV), and hard (2.0-4.5 keV) bands were used as an initial
measurement of their X-ray spectral properties. A hard X-ray spectrum or
hardness ratio (equivalent to a photon index of $\Gamma\sim$1) is indicative of a
pulsar binary (e.g.,Haberl \etal 2008). For the identification of
the optical counterparts of the \xmm sources we followed the analysis of Antoniou \etal
(2009b). We cross-correlated their coordinates with the OGLE-II
(Udalski \etal 1998) and MCPS (Zaritsky \etal 2002) catalogs, and 
searched for optical matches in a 5\arcsec\, radius around each
X-ray source (which includes the boresight error of \xmmsimple; e.g.,Brusa \etal 2007). Given the small number of X-ray sources with independently known optical counterparts, we cannot correct these observations for boresight errors. Based on the position of these
counterparts on the $V,B-V$ color-magnitude diagram, we identify
sources with early OB-type counterparts, while from hardness ratio
or spectral analysis we identified those hard X-ray sources ($\Gamma\sim$1), strongly suggesting they are XRB pulsars. Although
Monte-Carlo simulations indicate a significant number of spurious sources in
these fields, the identification of a hard X-ray source with an
early-type counterpart suggests that this is a true match.

We find that 15 \xmm sources have O- or B-type counterparts, while only
8 of those are hard X-ray sources, suggesting they are HMXBs.
Since all but one of the confirmed SMC HMXBs are Be-XRBs, they are also candidate Be-XRBs. Their properties are
presented in Table \ref{tableXMM}. The X-ray luminosity is derived assuming a
power-law spectrum of $\Gamma=1$ and \ion{H}{1} column density equal to 4.81,
6.63, and $4.51\times10^{20}$ $\rm{cm^{-2}}$ for fields 1,
2, and 3, respectively (based on Dickey \& Lockman
1990). The X-ray spectra of two sources with $>$200 counts were fitted with an absorbed power law, resulting in a photon index of $\Gamma=0.65\pm0.04$ and
$0.97\pm0.25$, and a column density of ${\rm
N_{H}}=(3.31\pm0.02)\times10^{20}$cm$^{-2}$ and $(0.30\pm0.15)\times10^{22}$cm$^{-2}$, for sources 2-1 (by simultaneously
fitting its MOS1 and MOS2 spectra) and 3-1 (from its PN spectrum), respectively. Source 2-1 in particular is a known
Be-XRB pulsar with a period of 169.3 s (Lochner \etal 1998) associated with emission-line object [MA93]623 (Meyssonnier \&
Azzopardi 1993; 3.3\arcsec\, away), source 3-1 remained unclassified in Sasaki \etal 2000 (ROSAT HRI src ID 11), while source 3-3 is the only one not included in the pipeline EPIC detection list.

If we include to the above sources the confirmed and candidate Be-XRBs
that lie within the \chandra and \xmm fields (from this work and those mentioned in \S \ref{intro}), we have a total of 54 (39) and 11 (2) HMXBs (Be-XRBs), respectively. For \chandra fields this is the sum of unique Be-XRBs, i.e., sources detected in two overlapping fields are counted once.

\section{SFH and XRB populations}\label{SFHfields}
The
recent SFH in our \chandra and \xmm fields is derived by averaging the spatially resolved SFH of the MCPS
regions ($\sim12\arcmin\times12$\arcmin; [HZ04]) encompassed by them. We find that:\\
\emph(i) For the \chandra fields, the most recent major burst peaked
$\sim$42 Myr ago, and it had a duration of $\sim$40 Myr. Moreover,
there were older SF episodes ($\sim$0.4 Gyr ago) with lower intensity
but longer duration, besides a more recent episode ($\sim$7 Myr)
observed only in \chandra field 4.\\
\emph(ii) For \xmm field 3, the most recent major burst occurred
$\sim$67 Myr ago. We also observed two fields with very young
populations (most recent major burst at $\sim$11 and 17 Myr ago
for fields 1 and 2, respectively). \xmm field 2 had an additional
intense burst $\sim$67 Myr ago (Table \ref{tableSFH}). 

In order to investigate the link between stellar and XRB populations,
we calculate the average SFH for the MCPS regions
($\sim12\arcmin\times12\arcmin$; [HZ04]) that host one or more Be-XRBs
(candidate and confirmed) detected in our \chandra and \xmm surveys
(39 and 2, respectively; see \S \ref{class}). The SFH in each region is weighted by the
encompassed number of Be-XRBs, and the error bars are derived based
on the upper and lower limits of [HZ04]. We repeat this exercise for the 15 MCPS regions without any known Be-XRB in our surveys. The two SFHs are presented in
Figure \ref{fig1}. The SFH of the Be-XRBs (black points) is strongly peaked at $\sim$42 Myr, while fields without any
Be-XRB (gray points) have minimal SFR at this age. This underscores
the difference between the fields with and without Be-XRBs, and suggests a clear connection between an SF event and the
observed Be-XRBs.

Following the above comparison, we also construct the SFH of the MCPS regions hosting one or more known
X-ray pulsars within any of our fields\footnotemark
\footnotetext{Based on the on-line census of Malcolm Coe
(http://www.astro.soton.ac.uk/$\sim$mjc/smc/ as of 2009 June 18).}
(Figure \ref{fig1}; black points), and for those that do not host such
sources (gray points). A
large fraction of these pulsars ($\sim$60\%; Liu \etal 2005) also appears in the Be-XRBs sample,
since the vast majority of their companions are Be
stars. This link is reinforced by the fact that all the counterparts of these X-ray sources lie on the region of the color-magnitude diagram consistent with main-sequence stars of age $\sim$40 Myr (Paper II). For completeness we present both, since the pulsar and the Be-XRB samples are selected on
the basis of their timing and optical properties, respectively. In total, in the MCPS regions that overlap with the \chandra fields lie 30 X-ray pulsars, while in the \xmm fields only 2 (sum of unique sources as in Section \ref{class}). As expected, the pattern in their SFH is very similar
to that of Be-XRBs. For regions rich in X-ray pulsars the SF peaks at
$\sim$42 Myr, while for regions without pulsars there is no peak at
this age.

The average SFH of the MCPS regions with and without any Be-XRBs detected in the \chandra Wing survey
(P.I. M. Coe; McGowan \etal 2008) is presented in Figure \ref{fig1}, top right
(black and gray points, respectively). This survey covered 20 fields
(3 of which are not used in this study because they do not overlap
with any MCPS region), and discovered 4 Be-XRB pulsars (Schurch \etal 2007). Repeating the same analysis, we find an SF peak at $\sim$42
Myr for fields with one or more known Be-XRBs. For regions in the Wing without Be-XRBs there is no SF burst at
this age; however, we see an intense burst at $\sim$11 Myr. For completeness, we
also present (Figure \ref{fig1}) the average SFH of
the MCPS regions with candidate (i.e., non spectroscopically confirmed) Be-XRBs from
the census of Liu \etal (2005), which also shows that they are
produced from the same SF burst as the pulsars and confirmed Be-XRBs. The above comparisons are summarized in
Table \ref{SFHpanels}.

The strong correlation between the number of XRBs and the age of the stellar populations at their location allows us to measure for the first time the XRB formation rate per unit SFR of their {\it parent} populations. The number of Be-XRBs (or HMXBs) per unit area detected in
our \chandra and \xmm surveys versus the SFR at $\sim$42 Myr for
the different fields is plotted in Figure \ref{fig2}. In order to have a homogenous
sample we used Be-XRBs detected only in these surveys. The best fit bisector line was calculated using the
``Linear Regression Software" (Akritas \& Bershady 1996), which takes
into account heteroscedastic errors. We find a slope of
$0.35\pm0.03$ Be-XRBs/SFR (or $0.40\pm0.04$ HMXBs/SFR), where SFR is in units of $10^{-3}$\msun/yr. This is the first direct
calibration of the XRB formation rate and the fact that it is based on
the source population in a single galaxy minimizes systematic effects
related to metallicity. For the same reason, this reflects the
Be-XRB formation rate for a low metallicity ($\sim1/5$\zsun; Luck
\etal 1998).

\section{Discussion}
\begin{figure}
\centering
\rotatebox{270}{\includegraphics[height=14.5cm]{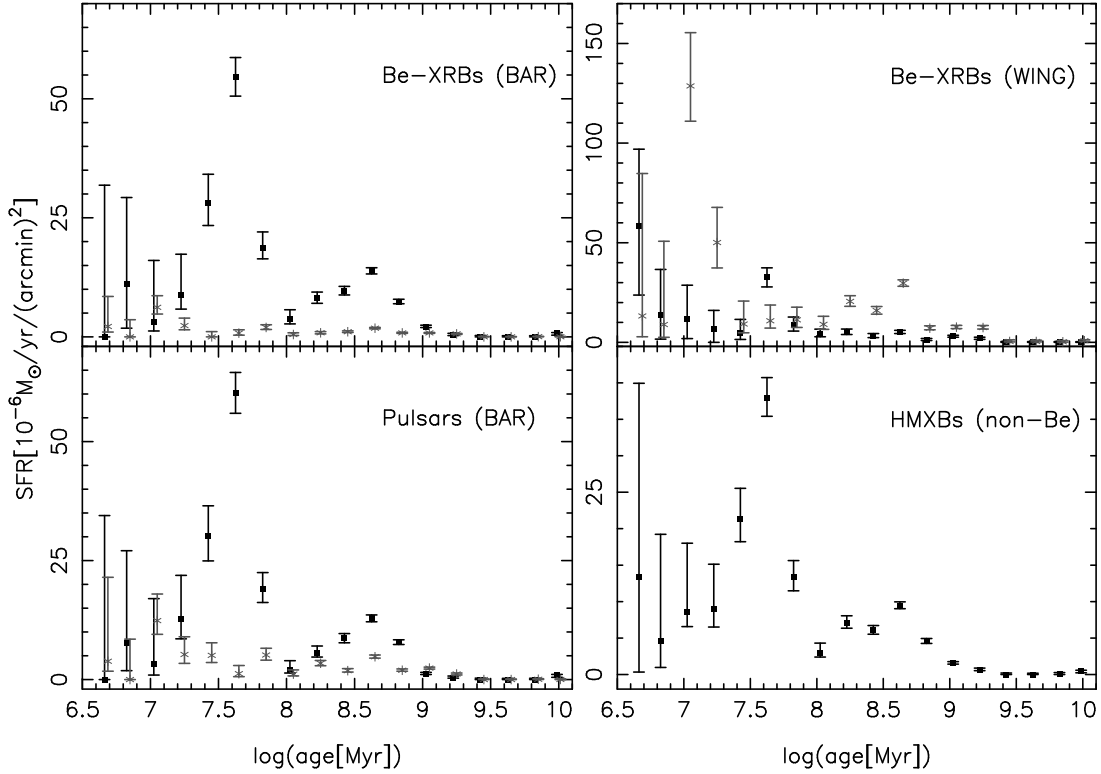}}
\caption{Average SFH (using data from [HZ04]) of
regions with and without XRBs (black and gray points, respectively) in
different locations of the SMC. {\it Top}: Be-XRBs in the Bar {\it (left)} and the Wing {\it (right)}. {\it Bottom}: pulsars in the Bar {\it (left)} and HMXBs over the SMC (excluding spectroscopically
confirmed Be-XRBs) from Liu \etal (2005; {\it right}). For clarity, a small offset of ${\rm
log(age[Myr])\sim0.025}$ has been applied on the distributions to
areas without Be-XRBs and/or pulsars.\label{fig1}}
\end{figure}

From the above analysis we find that the number of SMC XRBs peaks for
stellar populations of ages $\sim$25-60 Myr. In Figure \ref{fig1}, we also
see two additional peaks at $\sim$11 and $\sim$422 Myr. The one at
$\sim$11 Myr is too early to produce any pulsar XRBs, but could
result in a population of black-hole binaries (Belczynski \etal
2008) with O or early B-type donors which due to their massive companion evolve fast. The second SFR peak (at $\sim$422 Myr)
cannot result in HMXB formation, since by that time all OB stars have ended their lives.

The large number of Be-XRBs observed at ages $\sim$25-60 Myr is
consistent with the work of McSwain \& Gies (2005), who find that Be
stars develop their decretion disks at ages of $\sim$25-80 Myr, with a
peak at $\sim$40 Myr. OB stars formed $\sim$40 Myr ago are expected to reach the maximum rate of decretion disk formation at the current epoch.

A study of the evolution of XRBs by [SG07]
also found their maximum number at ages of 20-50 Myr after the SF
event, which however does not account for the Be phenomenon. They
interpret this peak in the HMXB numbers as the result of (1) the
pulsar spin-period evolution, (2) the nuclear evolution of the binary
system, and (3) the luminosity cutoff (${\rm
L_{X}}\sim10^{34}$\lunit) due to the sensitivity of the
observations. However, the luminosity cutoff
(e.g.,Linden \etal 2009) and evolution of the binary system may well
result in variations of the observed number of binaries at different
ages.

Another factor which may result in the excess of SMC 
HMXBs stems from the similarity between the epoch of the maximum occurrence
of Be stars and the ages of the stellar
populations hosting XRBs, and the fact their majority have Be-star donors. This indicates that the development of a decretion disk plays a major role in the overall statistics of the X-ray source populations by (1) increasing the number of active objects and (2) by increasing their observed luminosities due to the higher density and lower velocity of the outflow (Waters \etal 1988).

\begin{figure}
\centering
\rotatebox{270}{\includegraphics[height=12.cm]{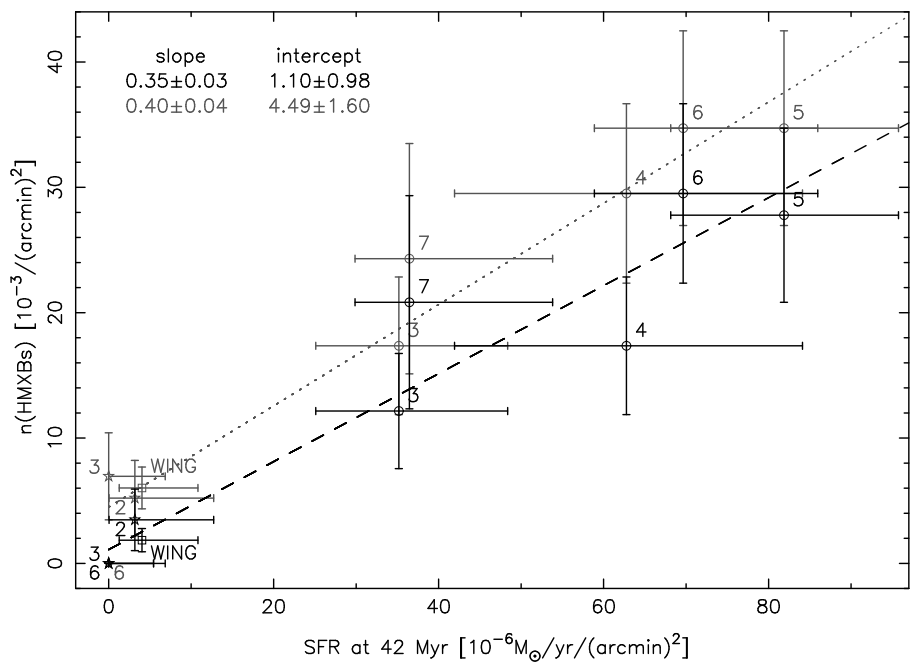}}
\caption{Number of observed Be-XRBs and HMXBs (shown in black and gray, respectively) in the \chandra and
\xmm fields vs. the SFR $\sim42$ Myr ago. \chandra (circles) and \xmm (asterisks) fields are marked with their IDs. The point marked as WING includes observations from \xmm field 1 and fields 5487, 5490, 5494 and 5495 from the \chandra Wing survey (P.I. M. Coe).\label{fig2}}
\end{figure}

This is also supported by the deficit of Be-XRBs in the SMC
Wing. Figure \ref{fig1} shows that the Wing has a weaker SF burst at the
age of enhanced formation of Be stars (i.e., at $\sim$42 Myr) than the
Bar, while its most recent SF burst occurred only $\sim$11
Myr ago. Thus, based on the above scenario, we do not expect a significant number of SMC Wing
Be-XRBs. Indeed, the number of observed sources is lower than that in
the SMC Bar, but consistent if we account for the SFR difference
at 42 Myr (Figure \ref{fig2}). On the other hand, an SF burst at these early ages ($\sim$11
Myr ago) suggests that supergiant HMXBs should dominate over Be-XRBs in the Wing. We also note that by comparing the number of binaries against the SFR (or the number of stars) in the same region
any projection effects cancel out.

The strong correlation between the number of XRBs and the
localized SF event can be used to constrain the kick velocity ($v_{kick}$) imparted
on the compact object during the SN explosion. In the case of large
kicks the XRBs would be scattered over
larger scales, diluting the correlation with their parent stellar
populations and resulting in lower contrast between the SFR of
regions with and without XRBs. Given an SF burst at $\sim$42 Myr and assuming a minimum pulsar birth
timescale of $\sim$10 Myr after the burst (e.g.,
Belczynski \etal 2008), the elapsed time since the kick is $\sim$30
Myr. In order to contain the XRBs within the spatial scale of the
star-forming regions ($\sim$40\arcmin; [HZ04]), we require a maximum
velocity of $\sim$15-20\kms. This is in agreement
with measured velocities of Be-XRBs in the Galaxy ($15\pm6$\kms;
van den Heuvel \etal 2000) and estimations derived from the mean distance
between a few pulsars and their nearest clusters in the SMC ($\sim$30\kms; Coe
2005b). Although these center of mass velocities are consistent with
typical SN kicks of $\sim$100\kms\, (Cordes \& Chernoff 1998), they could be much smaller given that the
XRBs show indication of local concentrations within the Bar associated with SFR enhancements in much smaller scales
($\sim$10\arcmin-15\arcmin; Paper II). This suggests at least a
factor of two lower $v_{kick}$ which would be consistent
with enhanced fraction of electron-capture SNe, which impart very low $v_{kick}$, as predicted
by Linden \etal (2009) for the SMC metallicity.\\

In this Letter, we discuss the importance of Be-XRBs as a
dominant component of young XRBs, based on a study of the
connection between X-ray source populations and their parent stellar
populations. We find that a significant number of Be-XRBs and/or
pulsars are connected with a burst of SF $\sim$25-60 Myr ago, while regions with weak SFR at $\sim$42
Myr, such as the SMC Wing, are deficient in Be-XRBs. We argue that the very strong similarity between the age of maximum
occurrence of Be stars and the age of the parent populations of XRBs
in the SMC indicates that the Be phenomenon plays a significant role
in the number of XRBs in this age range. Finally, based on the spatial correlation between the SF
activity and the XRBs, we set a limit on their $v_{kick}$ of
$\sim$15-20\kms\, while there is strong indication for velocities
of even a factor of two lower, and we estimate a Be-XRB production rate of $\sim$1 system per
$3\times10^{-3}$\msun/yr.

\acknowledgements
We thank Ewan O'Sullivan for helping with the \xmm data analysis, and the anonymous referee for comments which improved this Letter. This work was supported by NASA LTSA grant NAG5-13056, NASA grants G02-3117X, NNX08AB68G, NNG06GE68G, NNX10AH47G, and FP7 grant 206469 (ASTROSPACE).

\makeatletter
\def\jnl@aj{AJ}
\ifx\revtex@jnl\jnl@aj\let\tablebreak=\nl\fi
\makeatother
\ptlandscape 
\begin{deluxetable}{ccccccccc}
\tabletypesize{\scriptsize}
\tablecolumns{7}
\tablewidth{0pt}
\rotate 
\tablecaption{SFH and HMXB numbers\label{tableSFH}}
\tablehead{\multicolumn{3}{c}{Fields} & \multicolumn{3}{c}{Dominant SF Burst} & \multicolumn{3}{c}{Number}\\
\colhead{ID}  & \colhead{R.A.(J2000.0)} & \colhead{Dec.(J2000.0)} & \colhead{Age} & \colhead{Duration} & \colhead{SFR} & \colhead{HMXBs (Be-XRBs)} & \colhead{OB Stars} & \colhead{Pulsars}\\
\colhead{} &  \colhead{(h m s)} & \colhead{(\deg\, \arcmin\, \arcsec)} & \colhead{(Myr)} & \colhead{(Myr)} & \colhead{($10^{-6}$\msun/yr/arcmin$^{2}$)} & \colhead{} & \colhead{} & \colhead{}\\[0.2cm]
\colhead{[1]} & \colhead{[2]} & \colhead{[3]} & \colhead{[4]} & \colhead{[5]} & \colhead{[6]} & \colhead{[7]} & \colhead{[8]} & \colhead{[9]}}
\startdata
\chandra 3 & 00 56 46.14  &  -72 18 10.78  & 66.8    & 68      & $44.04_{-10.07}^{+10.07}$   &   10 (7) & 2220  & 6    \\[0.15cm]
\chandra 4 & 00 49 30.74  &  -73 16 52.34  & 42.2    & 36      & $62.76_{-20.83}^{+21.35}$   & 17 (10) & 4060  & 8    \\[0.15cm]
\chandra 5 & 00 53 11.45  &  -72 26 29.92  & 42.2    & 28      & $81.86_{-13.72}^{+13.89}$   & 20 (16) & 2730  & 12    \\[0.15cm]
\chandra 6 & 00 53 04.40  &  -72 42 18.22  & 42.2    & 36      & $69.64_{-10.76}^{+16.32}$   &  20 (17) & 3040  & 12    \\[0.15cm]
\chandra 7 & 00 49 25.00  &  -72 44 22.80  & 26.6    & 30      & $54.51_{-16.67}^{+25.35}$   &   7 (6) & 1670  & 3    \\[0.15cm]
\xmm 1     & 01 07 52.00 & -72 53 41.60 &  10.6   &   8     & $35.30_{-19.38}^{+26.88}$ & 4 (0)  & 3780                            & 0  \\[0.15cm]
\xmm 2     & 00 51 56.63 & -72 02 53.20 &  16.8   & 15      & $15.66_{-5.90}^{+17.83}$   & 3 (2)  & 3715  & 2  \\[0.15cm]
\xmm 3     & 00 42 25.45 & -73 36 29.40 &  66.8   & 39      & $15.65_{-6.83}^{+6.47}$ & 4 (0)  & 1500  & 0  \\[0.15cm]
\xmm 6     & 00 40 05.19 & -72 47 57.40 & 668.3   & $>$1200   & $4.35_{-0.99}^{+1.19}$  & 0 (0)  &  445      & 0  \\
\enddata
\tablecomments{Columns 1-3: observed fields (ID, R.A., Decl.); Columns 4-6: age, duration -defined as the
FWHM of its time evolution- and SFR; Columns 7-9: the number of HMXBs (Be-XRBs; see \S \ref{class}), OB stars (following
Antoniou \etal 2009b), and pulsars (see \S \ref{SFHfields}).}
\end{deluxetable}

\makeatletter 
\def\jnl@aj{AJ} 
\ifx\revtex@jnl\jnl@aj\let\tablebreak=\nl\fi 
\makeatother
\ptlandscape 
\begin{deluxetable}{ccccccccccc} 
\tablewidth{0pt} 
\tabletypesize{\scriptsize} 
\tablecolumns{0} 
\rotate 
\tablecaption{Properties of confirmed and candidate Be-XRBs detected with \xmm\label{tableXMM}} 
\tablehead{\colhead{Src} & \colhead{Source Name} & \colhead{R.A.(J2000.0)} & \colhead{Dec.(J2000.0)}  & \colhead{Net} & \colhead{S/N}  &    \colhead{${\rm L_{X}^{un}}$} &  \colhead{Optical}  & \colhead{Off. (Unc.)} & \colhead{{\it V}} &  \colhead{{\it B-V}} \\[0.1cm] 
\colhead{ID}  & \colhead{XMMU J} & \colhead{(h m s)} & \colhead{(\deg\, \arcmin\, \arcsec)} & \colhead{Counts} & \colhead{($\sigma$)}  & \colhead{($10^{34}$erg/s)}  & \colhead{src ID} & \colhead{($\arcsec$)} &  \colhead{(mag)}  & \colhead{(mag)}\\[0.1cm] 
\colhead{[1]} & \colhead{[2]} & \colhead{[3]} & \colhead{[4]} & \colhead{[5]} & \colhead{[6]} & \colhead{[7]} & \colhead{[8]} & \colhead{[9]} & \colhead{[10]} & \colhead{[11]}} 
\startdata 
 1-1\tablenotemark{\star}    & 010835.5-724308 & 01 08 35.54 & -72 43 08.4  & $63.87\pm11.11$ (1)                                      &  5.75 &               $2.79\pm0.52$  & O-11-104405                            & 3.82 (1.12)  & 17.82(2)      & -0.03(2)    \\
           &                                   &                       &                        &                                                                             &           &                                             & Z-4467654                                &  3.73 & 17.86(3)      & -0.04(4)    \\[0.1cm]
 1-2\tablenotemark{\star}    & 010519.9-724943 & 01 05 19.90 & -72 49 43.1 & $18.83\pm6.37$ (3)                                         &  2.96 &               $2.22\pm0.79$  & O-10-78741                               &  3.73 (1.43) & \nodata        & \nodata\tablenotemark{a}  \\
           &                                   &                       &                        &                                                                             &           &                                             & Z-4119599                                & 4.01  & 16.98(3)      & -0.09(4)    \\[0.1cm]
 1-3\tablenotemark{\star}    & 010620.0-724049 & 01 06 20.01 & -72 40 49.1  & $33.98\pm8.28$ (3)                                         & 4.10 &                 $5.45\pm1.42$ &  O-10-118866                           & 4.20 (1.59)  & 16.35(1)     & -0.02(2)  \\
           &                                  &                        &                        &                                                                             &           &                                             & $\equiv$O-11-13325              &             &                     &                      \\
           &                                  &                        &                        &                                                                             &           &                                             & Z-4232476                                &  4.53   & 16.38(3)     & 0.02(3)  \\[0.1cm]
 2-1    & 005255.1-715809 & 00 52 55.10 & -71 58 08.7 & $2155.48\pm53.72$ (2)                                  & 40.12 &             $135.75\pm3.40$ & Z-2430066                                &  2.82 (0.10)\tablenotemark{b}  & 15.53(2)     & -0.05(4)  \\[0.1cm] 
 2-2\tablenotemark{\star}    & 005149.3-720057 & 00 51 49.28 & -72 00 56.5  & $124.82\pm16.09$ (1)                                   & 7.76   &                $7.41\pm0.96$ &  Z-2274521                               & 3.46 (0.80)\tablenotemark{b}   & 18.38(3)     & -0.01(5) \\[0.1cm] 
 3-1    & 004208.0-734502 & 00 42 08.01 & -73 45 01.9 & $214.72\pm18.26$ (1)                                    & 11.76 &               $14.22\pm1.28$ & O-2-79541                                & 0.59 (0.46)   & 16.78(2)     & -0.05(5)  \\
           &                                   &                        &                       &                                                                             &            &                                             & Z-1132154                               & 0.40    & 16.78(4)    & -0.10(8) \\[0.1cm] 
 3-2\tablenotemark{\star}    & 004357.6-732840 & 00 43 57.57 & -73 28 39.7  & $24.56\pm6.34$ (2)                                        &   3.87  &               $2.63\pm0.70$ & O-3-122430                              & 0.65 (1.08)   & 18.31(2)    & -0.03(4)  \\
           &                                   &                        &                       &                                                                             &            &                                            & Z-1324298                                & 0.57    & 18.18(4)    & 0.12(5) \\[0.1cm] 
 3-3\tablenotemark{\star}    & 004514.7-733601 & 00 45 14.73 & -73 36 00.7  & $23.25\pm7.91$ (1)                                        &   2.94  &               $1.69\pm0.61$ &  O-3-178149                             & 2.73 (1.74)   & 15.17(1)    & -0.24(2) \\	   
           &                                   &                        &                       &                                                                            &             &                                            & Z-1466431                                & 2.62    & 15.19(10)  & -0.23(11)  \\
\enddata 
\tablecomments{Column 1: source ID as (field number)-(source ID in this field). Asterisks denote sources discovered within the \xmm observations presented here); Column 2: source name; Column 3: right ascension. Column 4: declination; Column 5: number of net source counts (0.2-12.0
keV; EPIC camera--1=PN; 2=MOS1; 3=MOS2); Column 6: source significance; Column 7: absorption corrected X-ray
luminosity (0.2-12.0 keV; see \S \ref{class}); Column 8: optical 
counterparts from OGLE-II as O-(field ID)-(source number) and MCPS as Z-(line number of the source in Table 1 of Zaritsky \etal 2002); Column 9:
distance between the counterpart and the X-ray
source (followed by the positional uncertainty of the X-ray sources given by the {\it edetect\_chain} task); Columns 10-11: apparent $V$-band
magnitude and $B-V$ color (with errors on the last significance digit).}
\tablenotetext{a}{Detected only in the $B$ and $I$-bands: $B$=16.87(1) mag, $I$=16.76(2) mag.}
\tablenotetext{b}{These offsets are within 2-3$\sigma$ above the positional uncertainty, but the (R.A.,Decl.) separations are in the same direction indicating a boresight error, which however, cannot be corrected (\S \ref{class}).}
\end{deluxetable}

\makeatletter
\def\jnl@aj{AJ}
\ifx\revtex@jnl\jnl@aj\let\tablebreak=\nl\fi
\makeatother
\begin{deluxetable}{cccccc}
\tabletypesize{\scriptsize}
\tablecolumns{6}
\tablewidth{0pt}
\tablecaption{SFH of different X-ray source populations\label{SFHpanels}}
\tablehead{\colhead{Region} & \colhead{Populations} & \multicolumn{2}{c}{Dominant SF Burst} & \multicolumn{2}{c}{SFR} \\
\colhead{} & \colhead{} &  \colhead{Age} & \colhead{Duration} & \colhead{of Most Intense Peak} & \colhead{at 42.2 Myr} \\
\colhead{} & \colhead{} &  \multicolumn{2}{c}{(Myr)} & \multicolumn{2}{c}{($10^{-6}$\msun/yr/(arcmin)$^{2}$)} \\[0.2cm]
\colhead{[1]} & \colhead{[2]} & \colhead{[3]} & \colhead{[4]} & \colhead{[5]} & \colhead{[6]}}
\startdata
SMC Bar & Be-XRBs                          & 42.2  & 33   & $54.50_{-3.94}^{+4.16}$        & $54.50_{-3.94}^{+4.16}$    \\[0.13cm]
                     & Non Be-XRBs                  & 10.6  &  7   & $6.22_{-1.44}^{+2.41}$           & $0.78_{-0.42}^{+0.71}$      \\[0.2cm]
SMC Bar & Pulsars                             & 42.2  &  33  & $60.09_{-4.14}^{+4.43}$        & $60.09_{-4.14}^{+4.43}$    \\[0.13cm]
                     & Non Pulsars                     & 10.6  &    5   & $12.40_{-2.90}^{+5.57}$        & $1.25_{-0.65}^{+1.70}$      \\[0.2cm]
SMC Wing  & Be-XRBs                          &  4.6   &   8   & $58.47_{-34.72}^{+38.48}$   & $32.73_{-4.91}^{+4.66}$    \\[0.13cm]
                     &                                            & 42.2\tablenotemark{a}  &  31   &      $32.73_{-4.91}^{+4.66}$                               &     $32.73_{-4.91}^{+4.66}$                  \\[0.13cm]
                     & Non Be-XRBs                  & 10.6  & 25   & $128.69_{-17.69}^{+26.73}$ & $10.85_{-3.69}^{+7.88}$    \\[0.2cm]
SMC Bar & Candidate Be-XRBs       & 42.2  & 31   & $37.86_{-2.45}^{+2.85}$        & $37.86_{-2.45}^{+2.85}$    \\[0.13cm]
\enddata
\tablecomments{Column 1: SMC region; Column 2: source
populations; Column 3: age of the most intense SF burst; Column 4: FWHM of burst's time evolution; Column 5: SFR of most intense peak; Column 6: SFR at 42
Myr.}
\tablenotetext{a}{Additional burst at ages $<$100 Myr.}
\end{deluxetable}

\end{document}